# Generation of bottle beam using low-density channel in air

Shao-jun Ji[1,2], Xiao-ming Zhou[1,2], Hui Tang[1,2], Hai-tao Wang[1,2,*], Jing-hui Zhang[1,2], Chun-hong qiao[1,2] and Cheng-yu Fan[1,2]
1 Key Laboratory of Atmospheric Optics, Anhui Institute of Optics and Fine Mechanics, Hefei Institutes of Physical Science，Chinese Academy of Sciences, Hefei, Anhui 230031, China
2 University of Science and Technology of China, Hefei, Anhui 230026, China
*Corresponding author: htwang@aiofm.ac.cn

**Abstract:** Cylindrical density depressions generated by femtosecond laser pulses filamenting in air for different energy depositions is investigated numerically, by using a set of hydrodynamic equations. The evolution of density profile is calculated for different temperature elevations, the results indicate that the gas density hole is getting shallower and wider with the increasing temperature elevations. A simulation of the propagation inside low-density channel implies a new way to generate a type of bottle beam.

## 1. Introduction

The filamentation of intense femtosecond laser pulses in gases is a special physical phenomenon in nonlinear optics, the complex physical process and wide applications contained in this phenomenon have attracted the interest of researchers all over the world, and the related theoretical and experimental studies are constantly emerging [1]. Filaments generated by femtosecond laser pulses in air are mainly a dynamic equilibrium process established between Kerr self-focusing effect and plasma defocusing effect induced by multi-photon ionization [2, 3]. Filamentation process is accompanied by a variety of radiation and pulse compression processes, such as super-continuous spectral radiation (SC) [4] and terahertz radiation (THz) [5, 6]. Moreover, the technological research on plasma filament as a new THz source [7], coherent X-ray source [8] and attosecond pulse source [9] is being continuously advanced. In addition, plasma filament can be used as a new virtual carrier to guide high voltage discharge [10], transmit electromagnetic wave [11] or video information [12]. A series of potential applications stimulate researchers to continuously explore the mechanism and controlling methods behind the filamentation of intense ultrashot laser pulses [13].

In the study of powerful femtosecond laser pulses filamentation, researchers have found that filamentation is not only a strongly nonlinear process of laser-matter interaction, but also a process of deposition of the laser energy in the medium that can stimulate the aerodynamics [14], which is expected to be developed into an air waveguide technology [15]. It is found that only a part of the laser energy is absorbed in the process of laser filamentation due to the ionization of air molecules and the inverse bremsstrahlung, which is only in the generation process of plasma channels during the filamentation. Besides ionization and molecular excitation, the rotational Raman absorption of air molecules will cause loss of laser energy. And this part of the laser intensity is lower than that required to generate plasma channel. However, after recombination of electron with ion and radiative decay of air molecules, this part of the ionization energy, in conjunction with the laser energy absorbed to ionize air molecules, will eventually translate into internal energy of air molecules, which is characterized by molecular heat [16]. After the filamentation, before the thermal diffusion of the air molecules, the energy deposition causes a change in the density of the air in the filament region, which propagates in the radial direction perpendicular to the propagation axis in the form of pressure waves (sound waves). And the resulting temporal evolution of gas number density or mass density has been investigated in early works [14, 16]. However, to our knowledge the evolution of the gas density hole for different temperature elevations caused by energy deposition is still unknown.

In this paper, we simulate the above hydrodynamic evolution in cylindrical geometry and investigate the depth and width of the gas density hole for different temperature elevations. In the second part, we use those results to calculate the refractive index variation in air. Lastly, we present simulations of the transmission of a Gaussian beam inside a 30-cm low-density channel.

## 2. Theoretical model

To gain an understanding of the underdense channel evolution for different energy deposition in air, the following set of equations for continuity, motion, and energy are solved numerically in cylindrical geometry using a one-dimensional finite difference method [17]:

$$\frac{d\rho}{dt} = -\rho \vec{\nabla} \cdot \vec{v}, \tag{1}$$

$$\rho \frac{dv_r}{dt} = -\frac{1}{m} \frac{\partial}{\partial r}(\rho k_B T), \tag{2}$$

$$\frac{5}{2} \frac{dT}{dt} = -T \vec{\nabla} \cdot \vec{v}, \tag{3}$$

where $k_B$ is the constant of Boltzmann, $v_r$ stands for the radial velocity, $m$ is the mass of air molecules, and $\rho$ is mass density. $T(r,t)$ is the neutral gas temperature profile, and we use the following radial profile as initial temperature perturbation:

$$T(r,0) = T_{bg} + dT \exp(-\frac{r^2}{\omega_0^2}), \tag{4}$$

where $T_{bg}$ is the background temperature (room temperature) of air, $dT$ is the temperature elevation achieved after repartitioning of the initial plasma energy to air and $\omega_0$ is the plasma channel waist.

In order to simulate the propagation of a Gaussian incident electric field inside low-density channel, a cylindrical scalar envelope, assumed to be slowly varying, is considered here, and it evolves according to the following paraxial scalar wave equation:

$$2ik_0 \frac{\partial E}{\partial z} + \nabla_\perp^2 E = 2k_0^2 (n_0 + \Delta n) E \tag{5}$$

where $k_0$ is the vacuum wavenumber. The term $\nabla_\perp^2$ accounting for the spatial diffraction denotes the transverse Laplacian operator, and $\Delta n$ represents the refractive index variation induced by acoustic gas dynamics.

## 3. Numeral results and discussions

An ideal gas equation of state is considered here. As the ambipolar diffusion of plasma resulted from femtosecond filamentation during the recombination process is much less than the original laser spot width, we adopt the value of $\omega_0 = 50\mu m$ [14] in our model and simulations. According to different experimental conditions, the temperature elevation due to energy deposition can range from about 100K [14] to more than 1000K [16], and it is reasonable to make it vary from 100K to 600K in our simulation.

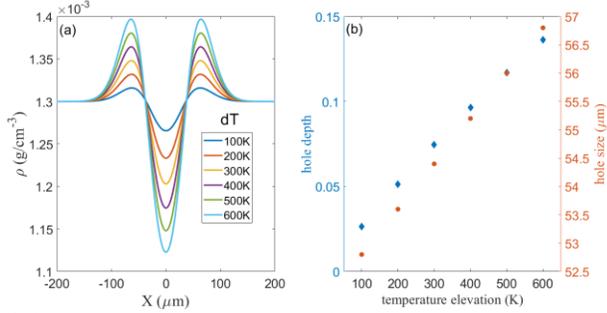

**Fig. 1.** (a) Evolution of the gas density hole for different energy deposition at the 50-ns pulse delay using hydrodynamic simulation. Initial conditions are $T_{bg} = 300K$ and $\rho_0 = 1.3 \times 10^{-3} g/cm^3$. From top to bottom, the temperature elevations owing to energy deposition of single femtosecond filament are 100K, 200K, 300K, 400K, 500K, and 600K, respectively.(b) Simulated evolution of depth (diamond) and width (dot) of the gas density hole vs. temperature elevation.

The hydrocode simulation of acoustic gas dynamics after femtosecond filamentation in air is shown in Fig.1 (a). From the radial distribution of air density, we know that the central density depression reaches its maximum depth when the temperature elevation is 600K. Moreover, the outward pressure waves enhance density in the region surrounding filament core, where the peak of density emerges.

In order to quantitatively investigate the hydrodynamic evolution for different energy deposition caused by femtosecond laser filament, the depth and width of the gas intensity hole are plotted in Fig.1 (b). The depth is defined by the ratio of the absolute value of on-axis reduction in gas density to the initial density $\rho_0$ and the width means the FWHM (full width at half maxima). It can be seen that the depth and width of gas intensity hole is getting deeper and wider with the increasing temperature elevation. Because the depth here is a dimensionless quantity, we employ the CV (Coefficient of Variation) to compare the effect of temperature elevation on depth with that on width. And its definition is the ratio of the standard deviation to the mean value. The datum in Fig.1 (b) give $CV_{depth} = 0.4912$ and $CV_{width} = 0.0273$. Apparently, the depth is more susceptible to the temperature elevation than the width.

In order to calculate the refractive index variation from the change in gas density, the relational expression $\Delta n/(n_0 - 1) = \Delta\rho/\rho_0$ was used, where $(n_0 - 1) = 2.8 \times 10^{-4}$ [18] is the refractive index of air at standard temperature and pressure. Fig. 2 shows that the higher the temperature elevation, the larger the shift in refractive index. The radial distribution of refractive index enables the formation of an annular waveguide structure.

In order to demonstrate the role of the annular waveguide, we simulate the propagation of a 532-nm beam in this kind of annular waveguide resulted from the acoustic waves generated by femtosecond filament, in the paraxial approximation by the means of BPM (Beam-propagation method) [19, 20]. The initial laser spot is shown in Fig. 3(a). In the absence of waveguide, the profile displayed in Fig. 3(b) shows that it still exhibits Gaussian shape after free propagation. The calculated intensity at different injected pulse delays for different temperature elevations is plotted in the lower panels of Fig. 3. As expected, the annular waveguide plays a role of spatial light modulator.

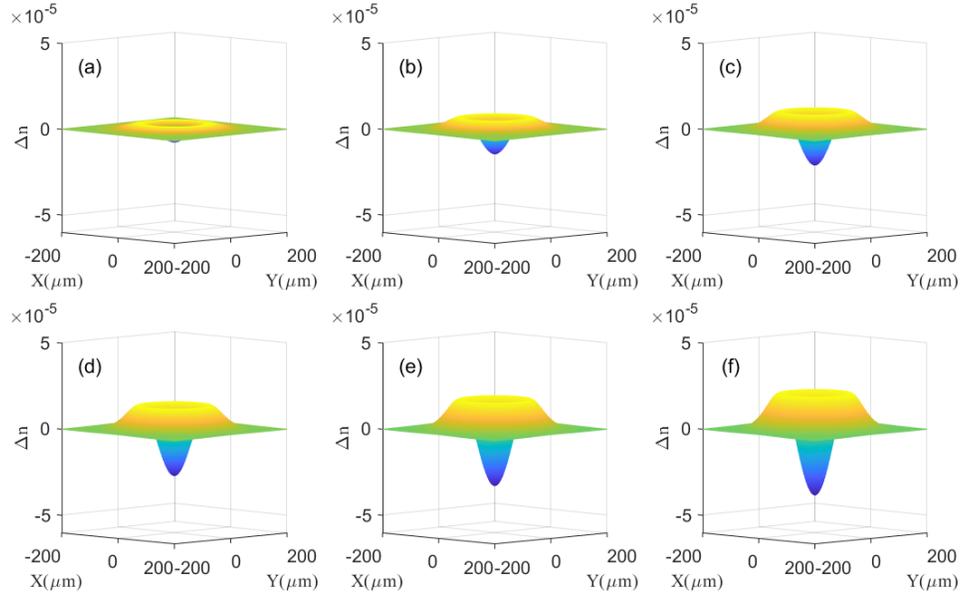

**Fig. 2.** Three-dimensional plots of the refractive index variation $\Delta n$ in air for different energy deposition, induced by acoustic gas dynamics. In panels (a)-(f), the $dT$ is 100K, 200K, 300K, 400K, 500K, and 600K, respectively.

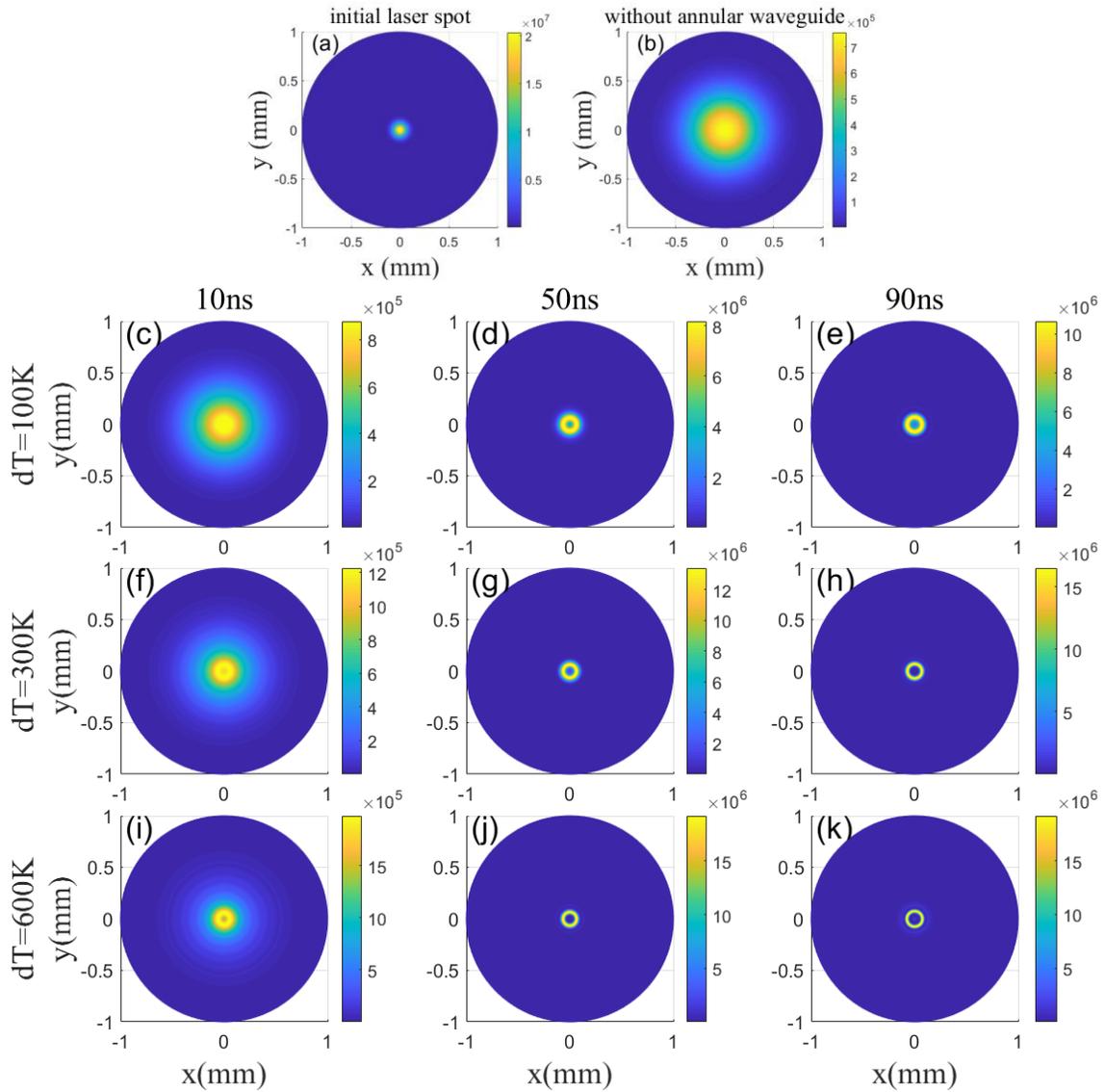

**Fig. 3.** The BPM simulation of the profile of the guided laser-beam at the output of a 30-cm annular waveguide structure.

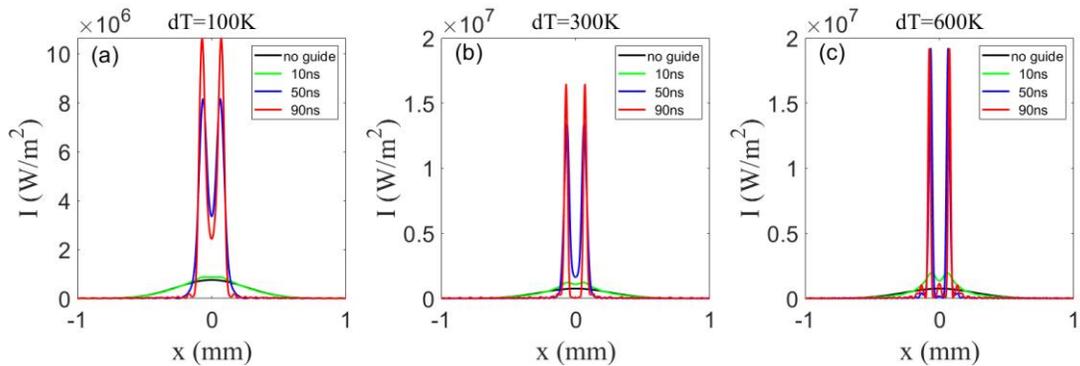

**Fig. 4.** The radial intensity distribution of the guided beam at the end of the annular waveguide structure.

To more clearly observe the intensity and spot size of the guided beam, we plot the radial distribution of intensity at different time delays in Fig. 4. By comparison, when the temperature elevation is 600K, the distribution at the 50-ns delay, namely the bottle beam shown in Fig. 3(j), is optimal within the scope of this paper, which is beneficial to choose appropriate conditions to obtain the bottle beam of desired shape and size.

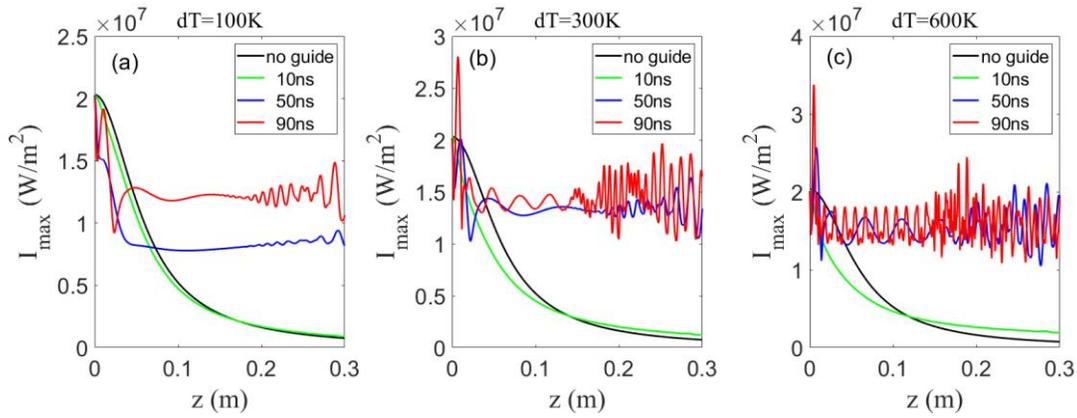

**Fig. 5.** The maximal intensity of the guided beam as a function of the propagation distance.

Finally, the evolution of peak intensity of the guided beam inside the annular waveguide structure is presented in Fig. 5. It shows that when the time delay is 10ns, the intensity smoothly decays even in the case where $dT$ is 600K, which indicates a week waveguide forms in early times after filamentation. However, the intensity is modulated high-frequently at longer delay for the higher temperature elevation. The reason for this phenomenon is that the impact of annular waveguide generated right now on beam is intense even if the spot size is small.

## 4. Conclusions

In summary, the evolution of low-density channel generated by acoustic shock wave after filamentation is investigated numerically for different energy depositions, by using hydrodynamic simulations in cylindrical geometry. Our results indicate that higher temperature elevation can result in larger on-axis density depression. The refractive index variation calculated from the change in gas density demonstrates the formation of an annular waveguide structure after filamentation in air. The radial intensity distribution of the beam guided by the above annular waveguide implies the emerging of an optimal bottle beam at 50-ns delay for the 600-K temperature elevation.

## Acknowledgments

This work was funded by the National Natural Science Foundation of China (Grant No. 61605223), and the Strategic Priority Research Program of Chinese Academy of Sciences (Nos. 306030105 and 614A010717). The project was also supported by Open Research Fund of State Key Laboratory of Pulsed Power Laser Technology (Nos. SKL2013KF01 and SKL2015KF03) and the dean Foundation of Hefei Institutes of Physical Science, Chinese Academy of Sciences (Grant No. YZJJ201506).